\documentclass{article}
\usepackage{amsmath,psfrag} 
 \usepackage{natbib}
 \usepackage{authblk}
 \usepackage{stmaryrd}
 \usepackage{graphicx}
  \usepackage[export]{adjustbox}
 \usepackage{siunitx}
\usepackage{float}
\bibpunct{[}{]}{;}{n}{}{,}
\def\rr#1{(\ref{#1})}

\DeclareMathOperator{\tr}{tr}

\newcommand{\be}{\begin{equation}}
\newcommand{\en}{\end{equation}}

\def\pafrac#1#2{\frac{\partial #1}{\partial #2}}
\def\dd#1#2{\frac{d #1}{d #2}}

\def\boldkappa{\mbox{\boldmath $\kappa$}}
\def\boldDelta{\mbox{\boldmath $\Delta$}}

\begin{document}
\title{Finite indentation of highly curved elastic shells}

\author[1,2]{\rm S.P.Pearce}
\author[3,4]{\rm J.R.King}
\author[5]{\rm T.Steinbrecher}
\author[5]{\rm G.Leubner-Metzger}
\author[6]{\rm N.M.Everitt}
\author[7]{\rm M.J.Holdsworth}

\affil[1]{School of Mathematics, University of Manchester, UK}
\affil[2]{Faculty of Biology, Medicine and Health, University of Manchester, UK}
\affil[3]{School of Mathematical Sciences, University of Nottingham, UK}
\affil[4]{Centre for Plant Integrative Biology, School of Biosciences, University of Nottingham, UK}
\affil[5]{School of Biological Sciences, Royal Holloway University of London, UK}
\affil[6]{Bioengineering Research Group, Faculty of Engineering, University of Nottingham, UK}
\affil[7]{Division of Plant and Crop Science, School of Biosciences, University of Nottingham, UK}
\renewcommand\Affilfont{\itshape\small}
\maketitle

\begin{abstract}
Experimentally measuring the elastic properties of thin biological surfaces is non-trivial, particularly when they are curved. One technique that may be used is the indentation of a thin sheet of material by a rigid indenter, whilst measuring the applied force and displacement. This gives immediate information on the fracture strength of the material (from the force required to puncture), but it is also theoretically possible to determine the elastic properties by comparing the resulting force-displacement curves with a mathematical model. Existing mathematical studies generally assume that the elastic surface is initially flat, which is often not the case for biological membranes. We previously outlined a theory for the indentation of curved isotropic, incompressible, hyperelastic membranes (with no bending stiffness) which breaks down for highly curved surfaces, as the entire membrane becomes wrinkled. Here we introduce the effect of bending stiffness, ensuring that energy is required to change the shell shape without stretching, and find that commonly neglected terms in the shell equilibrium equation must be included. The theory presented here allows for the estimation of shape- and size-independent elastic properties of highly curved surfaces via indentation experiments, and is particularly relevant for biological surfaces.
\end{abstract}

\section{Background}
\begin{figure}\centering
\includegraphics[width=0.9 \textwidth]{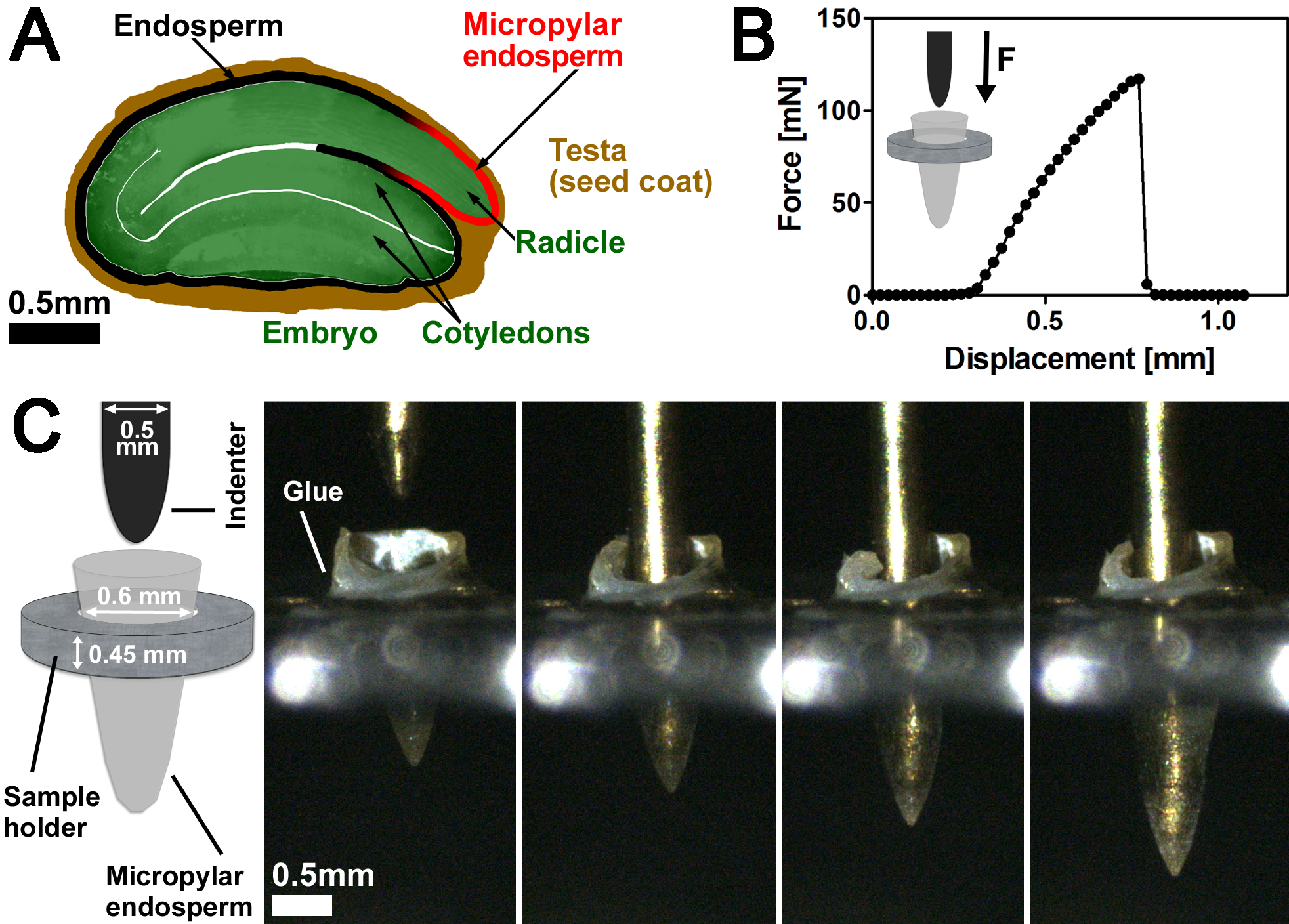}
\caption{(A) Sketch of a \textit{Lepidium sativum} seed, showing how the embryo (green) is enclosed by sections of the endosperm (black and red) with varying geometries. Image from \citet{muller2006}, Plant and Cell Physiology, 47(7), 864-877, used by permission of Oxford University Press. (B) Example force-displacement curve for an indentation of a Lepidium micropylar endosperm. (C) Sketch and corresponding still images showing how the endosperm stretches significantly during a typical indentation. Smaller diameter indenters are also used.}
\label{indentimage}
\end{figure}

The experimental characterisation of the elastic properties of a curved flexible shell is of interest within both biological and engineering contexts. Throughout biology, surfaces often grow with a three-dimensional structure, leading to complex curved shapes \citep{humphrey1998}. Such structures are not amenable to the majority of engineering techniques to determine elastic properties, such as vibration or tensile tests, since test-piece shapes must be controlled. Indentation tests are another classical technique, in which a rigid indenter is pushed into the specimen to generate a force-displacement curve (as shown in Figure \ref{indentimage}). With a suitable theoretical model, the elastic moduli of the sample can be extracted from such a curve, and this has been modelled for flat surfaces \citep{bhatia1968, yang1971, chudoba2000, begley2004, nguyen2004mech, komaragiri2005, steigmann2005, nadler2006}. However, little attention has been paid to curved surfaces at large indentation, with the context often that of atomic force microscopy (AFM) or nano-indentation, where the indentation depth and needle size are much smaller than the surface itself \citep{vella2011, vella2012}; \citet{deris2015}, who consider the indentation of a fluid filled spherical membrane being a recent exception. Without such a theoretical basis, shape-independent elastic properties can not be extracted from the experimentally measured force-displacement curves, and the sole readout is therefore force required before puncture, which gives information on the strength of the material but not the elasticity.

Additionally, the majority of the previous studies which have been conducted generally involve indentation from the convex side of the curved surface (the \lq outside', such as indenting a sphere, see Figure \ref{upsphRZindent}), assuming that the object is either internally pressurized or able to support its own weight, for example the studies by \citep{vaziri2008, vaziri2009, vella2011, vella2012, lazarus2012}. However, in biological samples this is often not the case, as the extracted tissues can be soft and not self-supporting, particularly under the action of an indenter. We therefore focus particularly on the case of indentation from the concave side (the \lq inside') of the surface (Figure \ref{sketches}), although the fundamental theory is applicable for the convex indentation too. 

Our particular motivation for considering this indentation problem are experiments performed on seeds of the Brassicaceae species \textit{Lepidium sativum} (garden cress) \citep{muller2006, lee2012, graeber2014}, in which the seed endosperm is punctured with a metal needle while the position and force are measured. Figure \ref{indentimage} shows images from such an experiment, showing the large indentation the endosperm can sustain as well as a sample force-displacement curve. This technique has been used in a wide range of different species, as detailed in Table 1 of \citet{steinbrecher2016}.

The Lepidium endosperm is non-uniform, with different sections of the embryo covered by portions of endosperm with different shapes (see Figure \ref{indentimage}), so the elasticity of these regions can not be compared without a model. As shown in Figure \ref{indentimage}, the endosperm is approximately a prolate spheroid with an aspect ratio between two and three, hence our interest in highly curved initial surfaces.

To fill this gap, we previously considered the indentation of a curved elastic membrane by a large rigid indenter \citep{pearce2011}, finding that such membranes are prone to wrinkling when indented, forming straight tension lines, such as those which may be seen when pushing into a plastic film (such as shrink-wrap) that is initially curved. This theory works well for moderately curved membranes (for instance oblate spheroids), but for more curved membranes (spheres or prolate spheroids) this theory gives deformations which are non-local, even for small indentation, this is due to the lack of bending stiffness meaning that no energy is required to bend the membrane without stretching. This means that when the initial shape of the membrane is sufficiently curved, no solution may be found for small indentation depths \citep{pearce2011}, as both of the principal stretches become compressive and the membrane becomes entirely slack, leading to a breakdown of the membrane theory. This is due to the membrane being able to freely change shape, pulling in the sides to be able to accommodate the additional stretch at the tip (see also Figure \ref{shellmembranecompare}). This behaviour is in contrast to that seen for indentation from the \lq outside' of a shell \citep{vaziri2008, vaziri2009} where wrinkling is seen only after a critical indentation distance as the initial deformation inverts a small section of material. 

In order to address these issues, we introduce here a shell theory instead of a membrane theory, regularising the singular problem by introducing higher order terms to provide bending resistance in the governing equations. In doing so we are able to find solutions for more curved shells, and solutions for small indentation distances generate only localised deformations of the shell. We find in particular that additional terms which are often neglected are required to ensure the force to indent the shell remains positive.

\section{Mathematical Formulation}
\subsection{Governing Equations}
\begin{figure}[!htb]
\psfrag{zz}{$R$}
\psfrag{yy}{$Z$}
\psfrag{xx}{$R(S)$}
\psfrag{ww}{\hspace{-0.4cm}$Z(S)$}
\psfrag{vv}{$S$}
\psfrag{uu}{$\Theta$}
\psfrag{aa}{$r$}
\psfrag{bb}{$z$}
\psfrag{cc}{$r(S)$}
\psfrag{dd}{\hspace{-0.5cm}$z(S)$}
\psfrag{ee}{$s$}
\psfrag{ff}{$\theta$}
\psfrag{gg}{$S_c$}
\psfrag{hh}{$s(S)$}
\psfrag{ii}{$\rho$}
\psfrag{jj}{$\phi$}
\psfrag{kk}{}
\includegraphics[scale=1,valign=t]{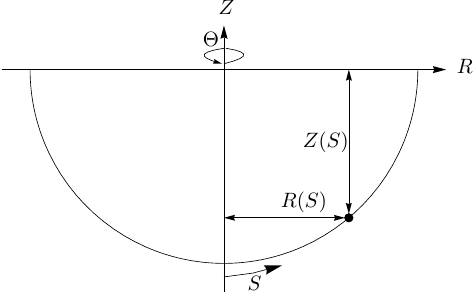}
\includegraphics[scale=1,valign=t]{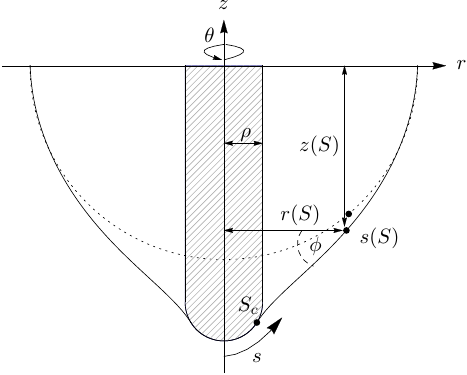}
\caption{Sketch of the two coordinate systems, the reference and deformed configurations.}
\label{sketches}
\end{figure}

We shall use the theory of nonlinear elasticity to model how a thin shell deforms under the action of an indenter, with the prescribed midsurface of the undeformed (or reference) shell defined parametrically by
\begin{equation}
\mathbf{X} = R(S) \mathbf{e}_R(\Theta) + Z(S) \mathbf{e}_Z, \quad 0 \leq S \leq L, \quad 0 \leq \Theta \leq 2 \pi,
\end{equation}
where $S$ shall be the independent variable throughout this work and we are using cylindrical coordinates. If the shell is initially spheroidal, we have 
\begin{equation}
R(S) = \sin S, Z(S) = - \gamma \cos S,
\end{equation}
where $\gamma$ is a parameter that controls the aspect ratio and the sign of $Z$ relative to the indenter controls the direction of indentation (with the negative sign showing indentation from the concave side), and we will use this throughout. In this case, $S$ is therefore the angle made with the axis of symmetry, and a flat shell is recovered for $\gamma=0$. While Figure \ref{sketches} shows an indentation from the concave side of the shell, this theory is also valid for indentation into the convex side, where $Z$ is positive. 

The origin of our coordinates lies directly under the indenter, at which point we require the smoothness conditions, $R(0)=0, Z'(0)=0$, where a prime denotes differentiation with respect to $S$. The system is invariant to rigid body movement in the $Z$-direction, so for convenience we choose $Z$ such that $Z(L)=0$. 
We consider deformations which map the reference configuration $\mathbf{X}$ onto the deformed configuration $\mathbf{x}$, the midsurface of which is given by
\begin{equation}
\mathbf{x} = r(S) \mathbf{e}_R(\theta) + z(S) \mathbf{e}_Z, \quad 0 \leq S \leq L, \quad 0 \leq \theta \leq 2 \pi,
\end{equation}
where we keep the same coordinate basis for simplicity. We assume that the initial configuration is axisymmetric so that the deformed configuration remains axisymmetric for an axisymmetric indenter, $\Theta = \theta$, and we impose the same smoothness conditions at the origin as in the reference configuration, $r(0)=0,z'(0)=0$. We solve the quasi-steady problem here, which is valid provided the indentation is sufficiently slow that inertial effects may be ignored, as it is in practice. As in \citep{pearce2011}, the principal stretch ratios may be defined in the tangential, azimuthal, and normal directions in the form
\begin{equation}\label{lambdas}
\lambda_s = \dd{s}{S} = \frac{\sqrt{\left(\dd{r}{S}\right)^2 + \left(\dd{z}{S}\right)^2}}{\sqrt{\left(\dd{R}{S}\right)^2 + \left(\dd{Z}{S}\right)^2}} \equiv \frac{\psi}{\Psi}, \quad \lambda_\theta = \frac{r}{R}, \quad \lambda_n = \frac{h}{H},
\end{equation}
where $h$ and $H$ are the deformed and undeformed thicknesses respectively, $s$ is the coordinate in the deformed configuration (this is not in general the arclength) and $\psi$, $\Psi$ are defined for convenience as the radicals in \rr{lambdas}$_1$. The principal curvatures, $\kappa_s, \kappa_\theta$, are defined by
\begin{equation}\label{curvatures}
\kappa_s =  \frac{1}{r'}\left(\frac{z'}{\psi}\right)' = - \frac{1}{z'}\left(\frac{r'}{\psi}\right)' , \; \kappa_\theta = \frac{z'}{r \psi},
\end{equation}
and curvatures in the reference configuration may be defined in an equivalent way, 
\begin{equation}\label{curvaturesRef}
\kappa^R_s =  \frac{1}{R'}\left(\frac{Z'}{\Psi}\right)' = - \frac{1}{Z'}\left(\frac{R'}{\Psi}\right)' , \; \kappa^R_\theta = \frac{Z'}{R \Psi}.
\end{equation}

The curvatures are related by Codazzi's equation,
\begin{equation}\label{codazzi2}
(r \kappa_\theta)' = r' \kappa_s,
\end{equation}
and the tangent, $\mathbf{e}_s= (r' \mathbf{e}_r + z' \mathbf{e}_z)/\psi$, and normal, $\mathbf{n}=(z' \mathbf{e}_r - r' \mathbf{e}_z)/\psi$, vectors are related by the Frenet-Serret equations,
\begin{equation}\label{frenet}
\mathbf{e_s}' = - \kappa_s \psi \mathbf{n}, \;\; \mathbf{n}'= \kappa_s \psi \mathbf{e}_s,
\end{equation}
both of which also hold for the reference configuration with appropriate substitutions.

The equilibrium equations for a membrane in the tangential and normal directions are respectively given by
\begin{subequations}\label{equib}
\begin{gather}
(r \tau_s)' -r' \tau_\theta =0, \label{equib1}\\
\kappa_s \tau_s + \kappa_\theta \tau_\theta = P,\label{equib2}
\end{gather}
\end{subequations}
where $P(S)$ represents the pressure difference across the membrane in the normal direction and $\tau_s, \tau_\theta$ are the principal stress resultants per unit length in the deformed shell (after integrating over the thickness) \citep{pearce2011}. This system \rr{equib} may be written as a third order system of ODEs in $\lambda_s,\lambda_\theta,\kappa_\theta$, or equivalently in $r,r',z'$, as $z$ does not appear explicitly.

When a thin-walled elastic shell is considered instead, by including bending moments, the governing equations which are commonly used are a generalisation of those in \rr{equib}, in our notation being written as
\begin{subequations}\label{equibshell}
\begin{gather}
(r \tau_s)' -r' \tau_\theta + \kappa_s r Q \psi =0, \label{equibsh1}\\
\kappa_s \tau_s + \kappa_\theta \tau_\theta - \frac{1}{r \psi} (r Q)' = P,\label{equibsh2}\\
(r M_s)' -r' M_\theta - r Q \psi =0, \label{equibsh3}
\end{gather}
\end{subequations}
where $M_s,M_\theta$ are the bending moments in the corresponding directions and $Q$ is the transverse stress resultant (shear stress) \citep{timoshenko1959, preston2008, pamplona2005, zarda1977}. The first two equations in \rr{equibshell} are the force balances in the tangential and normal directions, with the third equation being the balance of bending forces. The membrane limit \rr{equib} is given by \rr{equibshell} with $M_s, M_\theta$ both tending to zero. 

Equations \rr{equibshell} have been well studied by a number of authors \citep[for instance]{timoshenko1959, preston2008, pamplona2005, zarda1977}, generally assuming that the reference configuration is given by either a flat plate or spherical shell, and hence $\sqrt{R'^2 + Z'^2}=1$. They are also often written in Eulerian coordinates, with derivatives given as $\dd{}{s} = \psi^{-1} \dd{}{S}$, particularly when stretching is neglected entirely in favour of bending.   

We may integrate \rr{equibshell}, with the help of \rr{codazzi2}, to give the resultant force in the $Z$-direction as 
\begin{equation}\label{Flaw}
r^2 \kappa_\theta \tau_s - \frac{r r' Q}{\psi} = \int_0^S P r r' d S + \frac{F(S)}{2\pi},
\end{equation}
where we have defined the net axial force acting on the shell as $F(S)$, generated by the indenter. We will assume here that the pressure difference comes solely from the indenter, although the indentation of a pressurised membrane (such as a balloon or vesicle) may be incorporated into the framework presented here. As we are treating the indentation as a quasi-static process, $F$ is a function solely of $S$, with the experimentally measured force being $F(L)$. If the term involving the pressure in \rr{Flaw} is explicitly integrable, such as when $P$ is constant, we may use this equation to reduce the order of the system by one if we wish.
\subsection{Stress Resultants}\label{virtualwork}
Following \cite{steigmann1999,sanders1963,yeung1994}, the virtual work done by the shell may be written as:
\begin{equation}\label{dotU}
\dot{E} = \iint dA \left[ (T_s +  M_s \kappa_s) \frac{\dot{\lambda}_s}{\lambda_s} +(T_\theta + M_\theta \kappa_\theta  ) \frac{\dot{\lambda}_\theta}{\lambda_\theta} + M_s \dot{\kappa}_s + M_\theta \dot{\kappa}_\theta \right],
\end{equation}
where $T_s$ and $T_\theta$ are planar stress resultants and dots indicate the variation of a quantity. This expression may be derived from three-dimensional theory by integration through the thickness \cite{sanders1963,yeung1994}, assuming the Kirchhoff hypotheses.

We can thus see that the moment resultants are conjugate to the virtual changes in curvature, but the terms conjugate to the virtual changes in the stretches are \lq generalised tensions' \cite{yeung1994}, which take the form of tension plus curvature times bending moment. It is these generalised tension stress resultants that feature in \rr{equibshell}, and they therefore should be of this generalised form, including the $M \kappa$ terms, and so we will define
\begin{equation}\label{tensions}
\tau_s = T_s + \xi M_s \kappa_s , \;\; \tau_\theta = T_\theta + \xi M_\theta \kappa_\theta,
\end{equation}
where the parameter $\xi \in \{0,1\}$ is zero for the commonly used \lq first approximation' \cite{pamplona2005} and one for the full expansion as required by \rr{dotU}. These additional terms arise from the fact that a small shell element is actually curved rather than flat as implicitly assumed in \rr{equibshell}, and so the tension and bending moments are coupled together due to rotations of the surface; see \cite{pamplona2005} for more details. We will show here that when the shell is highly curved these terms become significant and cannot be ignored; an explanation in Cartesian coordinates is found in Article 328 of \cite{love1927}. These additional terms appear naturally in the theory of \cite{steigmann1999}, as well as others \cite{sanders1963,pietraszkiewicz1989,pietraszkiewicz1993,opoka2009} which derive the governing equations from a variational principle as opposed to a force balance. The necessity of these terms in the buckling of liposomes is discussed in \cite{pamplona2005}. Similar terms are used in \cite{blyth2004, evans1994}, but with the opposing curvatures in the additional terms, i.e. $\tau_s = T_s + M_s \kappa_\theta$; this may have come from the presence of these opposing curvatures in the definition of the integrals in the three-dimensional theory, the reason for this discrepancy is not clear.

\subsection{Constitutive Equations}\label{conseqns}
It remains to specify the constitutive equations relating the stresses and bending moments to the stretches and curvatures. As for the membrane case we suppose the existence of a strain-energy function $W(\lambda_s,\lambda_\theta, \lambda_n)$, and assume that the material is incompressible and therefore set $\lambda_n = \lambda_s^{-1} \lambda_\theta^{-1}$ \citep{ogden1997}. The principal stress resultants per unit length in the deformed shell are then defined from $T_\alpha = h \sigma_\alpha, \alpha \in \{s,\theta\}$, where $\sigma_\alpha = \lambda_\alpha \pafrac{W}{\lambda_\alpha}$ is the usual principal Cauchy stress in incompressible three-dimensional elasticity, leading to 
\begin{equation}\label{stressres}
T_s = \frac{H}{\lambda_\theta} \pafrac{W}{\lambda_s}, \quad
T_\theta = \frac{H}{\lambda_s} \pafrac{W}{\lambda_\theta},
\end{equation}
for further details see \citet{naghdi1977,haughton2001,pearce2011}. While the theory presented here is appropriate for any isotropic incompressible strain-energy function, we will mostly show examples with the Mooney-Rivlin material model,
\begin{equation*}
W^{MR} = \frac{\mu}{2}\left[(1-\alpha) \left( \lambda_s^2 + \lambda_\theta^2 + \lambda_s^{-2} \lambda_\theta^{-2} - 3 \right) + \alpha \left( \lambda_s^{-2} + \lambda_\theta^{-2} + \lambda_s^{2} \lambda_\theta^{2} - 3 \right) \right], 
\end{equation*}
where $\mu$ is the shear modulus and $\alpha$ controls the deviation from the Hookean response (the neo-Hookean strain-energy function being given by $\alpha = 0$). We will also compare with a strain-stiffening  Gent-type model \citep{gent1996}, 
\begin{equation*}
W^{G} = \frac{-\mu}{2} J_m \log(1-  \frac{(\lambda_s^2 + \lambda_\theta^2 + \lambda_s^{-2} \lambda_\theta^{-2}) - 3}{J_m}),
\end{equation*}
where $J_m$ is a positive parameter representing a maximum value beyond which the hydrocarbon chains can not stretch any further. Rubbers are commonly described with values of $J_m$ of $97.2$ or $114$ \citep{gent1996}, but for stiff biological tissues values as small as $0.4$ have been used \citep{horgan2015}.
However, as we also want to include the dependence of the energy on the effect of bending, it is necessary also to give a constitutive equation for the bending moments. Appropriate forms for these dependencies are not clear in the literature, with various assumptions often being made without any clear justification or only applying in specific cases, such as area conserving deformations. Here we shall follow the derivation of \citet{steigmann1999}, where the higher order bending effects are dependent only on $\nabla \mathbf{F}$, where $\mathbf{F} = \pafrac{\mathbf{x}}{\mathbf{X}}$ is the surface deformation gradient, thereby incorporating only the elastic resistance to flexure in addition to the standard strain resistance. This leads to an energy $U(\mathbf{C},\boldDelta;\mathbf{X})$, where $\mathbf{C} = \mathbf{F}^T \mathbf{F}$ is the right Cauchy-Green tensor and $\boldDelta = \boldkappa - \boldkappa^R$ is the relative curvature strain tensor (see \citep{steigmann1999} for details). The first two invariants of the relative curvature strain tensor are given by
\begin{align}
\tr \boldDelta& = (\lambda_s^2 \kappa_s - \kappa_s^R) + (\lambda_\theta^2 \kappa_\theta- \kappa_\theta^R),\\
\det \boldDelta &= (\lambda_s^2 \kappa_s -\kappa_s^R) (\lambda_\theta^2 \kappa_\theta - \kappa_\theta^R),
\end{align} 
and are related to the mean and Gaussian curvatures respectively. There are three further invariants which involve the coupling between $\mathbf{C}$ and $\boldDelta$, but we do not consider them here. We explicitly note that all the invariants of $\boldDelta$ involve both the stretches and the curvatures, which means that the bending moments should be based on these kind of mixed terms when bending and stretching is occurring, not just relative curvature changes. 
As the shell is isotropic, the energy must be invariant under the rotation of the coordinate system and hence be an even function of $\tr \boldDelta$, so the simplest appropriate form for the energy $U$ is
\begin{equation}\label{Uenergy}
U =  W(\lambda_s,\lambda_\theta) + \frac{B}{2} (\tr \boldDelta)^2,
\end{equation}
where $B$ is a bending modulus and we choose not to involve the Gaussian curvature related term $\det \boldDelta$. The bending moments are thus given by
\begin{subequations}
\begin{gather}\label{momentconst}
M_s = h \pafrac{U}{\kappa_s} = B H \frac{\lambda_s}{\lambda_\theta} (\lambda_s^2 \kappa_s + \lambda_\theta^2 \kappa_\theta - \kappa_s^R - \kappa_\theta^R), \\
M_\theta = h \pafrac{U}{\kappa_\theta}=B H \frac{\lambda_\theta}
{\lambda_s} (\lambda_s^2 \kappa_s + \lambda_\theta^2 \kappa_\theta- \kappa_s^R - \kappa_\theta^R).
\end{gather}
\end{subequations}
Other constitutive equations for combined bending and stretching have been used within the literature \citep{pamplona2005, blyth2004, preston2008} but do not involve the invariants as presented above, being based on more ad-hoc assumptions. For incompressible linear elasticity, the bending modulus $B$ is proportional to the shear modulus \citep{timoshenko1959},
\begin{equation}\label{Bdef}
B = \frac{\mu H^2}{12},
\end{equation}
and we note that the third power of $H$ is already included in the definition of the bending moments \rr{momentconst}, as we are working in terms of integrated stress resultants.

\section{Solution Procedure}
When written in terms of $r$ and $z$, \rr{equibshell} appears to be a seventh order ODE system (as $z$ never appears undifferentiated). However, the highest order derivatives appear only in specific combinations, so it is actually a set of five nonlinear first order ODEs in $\lambda_s,\lambda_\theta,\kappa_s,\kappa_\theta,Q$, with \rr{codazzi2} and \rr{lambdas}$_1$ providing two further equations, after which $z$ may be found by integrating \rr{curvatures}$_2$.

To avoid square roots in the numerical calculations, we introduce $\phi$, the angle between the axis of revolution and the normal to the meridian in the deformed configuration (see Figure \ref{sketches}), defined by                                                                                                                                                                                         
\begin{equation}
\frac{d r}{d s} = \frac{r'}{\psi} = \cos \phi,\;\; \frac{d z}{d s} = \frac{z'}{\psi} = \sin \phi, \;\; \kappa_\theta = \frac{\sin \phi}{r}, \;\; \kappa_s = \frac{\phi'}{\psi}.
\end{equation}
The undeformed domain will be split into two regions, based on whether their corresponding material points are in contact with the indenter or not, with the boundary circle being given by $S = S_c$. This contact circle is unknown \textit{a priori}, and must be determined as part of the solution, and so we therefore need to give six boundary conditions to close the fifth-order system.
\subsection{Indenter Region}
We assume the shell is indented by a rigid indenter consisting of a cylinder with radius $\rho$ connected to a tip which is described parametrically in terms of an angle $\omega$ by $r=\rho A(\omega), z=\rho B(\omega)$, where $A$ and $B$ are specified.  We require the axisymmetry requirements $A(0)=0, B'(0)=0$ at the axis, and hence only consider smooth indenter tips here. Indentation by an isolated sphere or other axisymmetric object under gravity, for example, could also be accommodated in the same framework.
In the contact region, we assume that the shell conforms precisely, prescribing both $r$ and $z$ there, although the stretch in this contact region is still unknown; we assume there is no slip between the indenter and the shell. This may be violated in the case of buckling, particularly when compressive stresses occur, but we leave this as future work. We therefore let
\begin{equation}\label{indentrzdef}
r(S) = \rho A(\omega(S)), \; z(S) = - \delta + \rho B(\omega(S)),
\end{equation}
where $\delta$ is the depth of indentation, to be found as part of the solution, and we treat the angle $\omega$ in the deformed configuration as a function of $S$. Therefore in the contact region,
\begin{equation}\label{indentstretch}
\lambda_s = \rho \frac{\sqrt{A'(\omega(S))^2 + B'(\omega(S))^2}}{\Psi} \dd{\omega}{S}, \; \lambda_\theta = \rho \frac{A(\omega(S))}{R(S)},
\end{equation}
where we have used $\omega'(S) >0$. We may then evaluate the first equilibrium equation, \rr{equibsh1}, to find $\omega(S)$ given appropriate boundary conditions, having solved the third equilibrium equation \rr{equibsh3} for $Q$. The second equilibrium equation \rr{equibsh2} enables us to calculate the pressure $P$ exerted by the indenter on the shell after calculating the deformation (and hence the function $F$), but this is supplemental to computing the deformation itself in the contact region. Equation \rr{equibsh1} leads to a second-order differential equation for $\omega(S)$, as the principal curvatures become
\begin{align}\label{indentercurvatures}
\kappa_s&= \frac{A'(\omega) B''(\omega) - B'(\omega) A''(\omega)}{\rho \;\omega'(S) (A'(\omega)^2 + B'(\omega)^2)^{3/2}}, \\ 
\kappa_\theta&= \frac{B'(\omega)}{\rho \;\omega'(S) A(\omega)\sqrt{A'(\omega)^2 + B'(\omega)^2}},
\end{align}
and so $\kappa_s$ involves $\omega'(S)$ but not $\omega''(S)$ as might have been expected. This means that the governing equation in the indenter region is second order in $\omega(S)$, as in the membrane case.

At the pole, $S=0$, there exists a coordinate-induced singularity in the governing equations, as $r$ and $R$ are both zero at this point, as is also true in the membrane case. We therefore begin the integration in the indenter region at a value $0<\zeta \ll 1$, at which we use the expansions
\begin{equation}\label{deltaexpand}
\omega(\zeta) = \zeta \omega'(0) + \mathcal{O}(\zeta^3), \quad \omega'(\zeta) = \omega'(0) + \mathcal{O}(\zeta^2),
\end{equation}
where we have used the fact that $\omega$ is an odd function and $\omega(0)=0$, one of the boundary conditions. Using the restrictions on $A$ and $B$, we find $\omega'(0) = \rho^{-1} \lambda_0 R'(0)/A'(0)$, where $\lambda_0 \equiv \lambda_s(0) = \lambda_\theta(0)$ is the value of the stretch at the pole, used in a shooting procedure. So our boundary conditions at the pole of the contact region are,
\begin{equation}
\omega(\zeta) = \zeta \rho^{-1} \lambda_0 R'(0)/A'(0), \quad \omega'(\zeta) = \rho^{-1} \lambda_0 R'(0)/A'(0),
\end{equation} 
and we can integrate to $S=S_{cyl}$ where the shell contacts the cylindrical part of the indenter, for a given $\lambda_0$. Cases of extreme indentation where the cylindrical part of the indenter touches the shell may also be calculated, with appropriate changes to \rr{indentrzdef}. Varying $\lambda_0$ allows us to change the indentation depth, $\delta$, and we note that for small indentations for curved shells there may be a compressive stretch at the pole, with $\lambda_0<1$.

This method assumes that the indenter first comes into contact with the tip of the shell, and will not work if the side of the shell is touched first, hence we require the curvature of the indenter to be less than that of the shell, $\kappa_\theta^R(0) < B''(0)/(\rho A'(0)^2)$.
\subsection{Free Region}
Outside of the contact region, the equations \rr{equibshell} are fifth order, with $P$ being zero as there is no applied pressure in this region. In addition, as we do not know the location of the contact circle, $S_c$, at which the continuity conditions will be specified, we require six boundary conditions to complete the system. At the fixed boundary $S=L$ we need to apply appropriate boundary conditions, here we assume that the shell is simply supported (hinged) at a fixed radius, and thus $\lambda_\theta = 1, M_s = 0$. This is appropriate for the experimental setup shown in Figure \ref{indentimage}, where the seed endosperm is slotted into a cylindrical hole. 

We note that a radial pre-stretch prior to the indentation could be included, by allowing $\lambda_\theta(L) = \lambda_p>1$, but this pre-stretch will necessarily change the shape of the shell prior to indentation and induce significant additional complexity when the surface is not initially flat. This may be relevant when a non-zero internal pressure is included or to account for growth in biological contexts.

At the contact circle, we require continuity of the position, normal and the resultant force, $F$, giving four continuity conditions. When using the approximate expansion ($\xi=0$), this implies that $\lambda_s, \lambda_\theta$, $\kappa_\theta$ and $Q$ are continuous. The remaining variable, the curvature $\kappa_s$, is allowed to have a jump at the interface, as in the membrane case \citep{pearce2011}, enabling the shell to change curvature between the forced indenter shape and the remaining free section. 
However, when using the full expansion $\xi = 1$, the force $F$ now includes $\kappa_s$, and this then requires $\lambda_s$ to be discontinuous via a jump in $\psi$, while keeping $\phi$ continuous to ensure continuity in the normal. This ensures that both generalised tensions are continuous across the interface, and we therefore have
\begin{subequations}\label{jumpeqns2}
\begin{gather}
\llbracket \lambda_\theta \rrbracket = \llbracket \kappa_\theta \rrbracket = \llbracket Q \rrbracket = 0, \;\; \llbracket F \rrbracket = \llbracket r \sin \phi \; \tau_s - r \cos \phi \; Q \rrbracket = 0 \label{jumpF}\\
\lambda_\theta(L) = 1, \;\; M_s(L) = 0,
\end{gather}
\end{subequations}
where $\llbracket x \rrbracket = x(S_c^{+})- x(S_c^{-})$ is the jump in the value of $x$ across the contact line; these conditions imply that $\llbracket \phi \rrbracket = 0$. It is most convenient numerically to solve \rr{jumpF} for $\lambda_s(S_c^{+})$, given $\kappa_s(S_c^{+})$. We then vary both $S_c$ and $\kappa_s(S_c^{+})$ in order to satisfy the two boundary conditions at $S=L$, using a shooting method.

Given the initial size and shape of the indenter and shell, as well as constitutive equations for the shell, we can solve the system above for a specified $\lambda_0$. We then calculate the force on the indenter, $F$, and the depth of indentation of the shell, $\delta$, and then vary $\lambda_0$ to cover the range of forces and displacements, allowing to the construction of a force-displacement curve.
\subsection{Non-dimensionalisation}
All the previous expressions involve dimensional quantities, which we will now denote with hats in this section. We now non-dimensionalise by choosing the undeformed radius $\hat{R}_L = \hat{R}(\hat{L})$ as our characteristic length, and we have the choice of either $\hat{\mu} \hat{R}_L \hat{H}$ or $\hat{B}$ as our characteristic force. To be consistent with \citet{pearce2011} and in order to keep the membrane limit approachable without rescaling, we choose $\hat{\mu} \hat{R}_L \hat{H}$. Therefore we have
\begin{gather}
(\hat{S},\hat{R},\hat{Z},\hat{H},\hat{r},\hat{z},\hat{\rho},\hat{\delta},\hat{L}) = \hat{R}_L (S,R,Z,H,r,z,\rho,\delta,L) \\
(\hat{T}_s,\hat{T}_\theta,\hat{Q}) = \hat{\mu} \hat{H} (T_s,T_{\theta},Q), \hat{P} = \hat{\mu} P, \;\;
\hat{F} = \hat{\mu} \hat{R}_L \hat{H} F, \;\; \hat{W} = \hat{\mu} W, (\hat{M}_s,\hat{M}_\theta) = \hat{\mu} \hat{R}_L \hat{H} (M_s,M_\theta).\nonumber
\end{gather}
Having done this non-dimensionalisation, the governing equations \rr{equibshell} remain the same, except for $\hat{P}$ being replaced by $P/\epsilon$, where $\epsilon = \hat{H}/\hat{R}_L$. In the indenter region $\rho$ becomes the ratio of the radii of the indenter and the initial membrane. The constitutive equations \rr{momentconst} become
\begin{gather}
T_s = \frac{1}{\lambda_\theta} \pafrac{W}{\lambda_s}, \;\; M_s = \beta \frac
{\lambda_s}{\lambda_\theta} (\lambda_s^2 \kappa_s + \lambda_\theta^2 \kappa_\theta- \kappa_s^R - \kappa_\theta^R) \\
T_\theta = \frac{1}{\lambda_s} \pafrac{W}{\lambda_\theta},  \;\; M_\theta = \beta \frac{\lambda_\theta}{\lambda_s} (\lambda_s^2 \kappa_s + \lambda_\theta^2 \kappa_\theta- \kappa_s^R - \kappa_\theta^R),\nonumber
\end{gather}
where the dimensionless parameter, $\beta \equiv \frac{\hat{B}}{\hat{\mu} \hat{R}_L^2}$ describes the relative contribution of bending to stretching. 
If we assume the linear elasticity relationship \rr{Bdef} holds we also have 
\begin{equation}
\beta = \frac{\hat{H}^2}{12 \hat{R}_L^2} = \frac{\epsilon^2}{12}.
\end{equation}
\subsection{Numerical Solution}
To recap, we solve equation \rr{equibsh1} in the contact region with the substitutions \rr{indentrzdef} for a given $\lambda_0$, to find a solution in which the indenter remains in contact with the shell for a range of $S$. This may be done for any $\beta$, as the highest derivatives in these equations are not solely multiplied by $\beta$.

We then solve the equilibrium equations \rr{equibshell} in the free region, adopting a shooting approach on the values of both $S_c$ and $\kappa_s(S_c)$, and integrate through the free region to $S=L$. 

After this we iterate on $S_c$ and $\kappa_s(S_c)$ to satisfy the remaining two boundary conditions at $S=L$, to find the indentation distance $\delta$ and force resultant $F$ for the prescribed $\lambda_0$. Varying $\lambda_0$ then allows us to generate a set of solutions with different indentation distances, allowing us to characterise the entire indentation process. It should be noted during the integration process, if the initial guesses for $\kappa_s(S_c)$ and $S_c$ are too incorrect, the system often can become stiff or even singular when the deformed radius vanishes, especially when the full expansion is used and/or $\beta$ is small (as in the free region this is a singular perturbation of the membrane case). The use of continuation of the values from neighbouring values of $\lambda_0$ therefore greatly helps the convergence of this numerical procedure. All calculations were undertaken using Mathematica 10 \citep{mathematica2014}, code available on request.
\section{Results}
Previously, as detailed in \citet{pearce2011}, we used the tension-field membrane theory in order to model the indentation of a membrane. With this theory, when a stress becomes compressive it is assumed that the membrane wrinkles, forming non-axisymmetric tension-lines perpendicular to the compressive stress direction, but a smoothed pseudo-surface can be calculated. As discussed in \citet{pearce2011}, when the reference surface becomes significantly curved ($\gamma > 0.5$), the tension-field membrane theory breaks down for $\lambda_0$ close to unity (and hence small $\delta$), and no solution is able to be found as both stresses become negative and the membrane is entirely slack. This occurs because, without the inclusion of bending resistance, the membrane can pull in throughout the free region with minimal stretching.

\begin{figure}[H]
\psfrag{aa}{$r$}
\psfrag{bb}{$z$}
\includegraphics[height=2.5in,keepaspectratio]{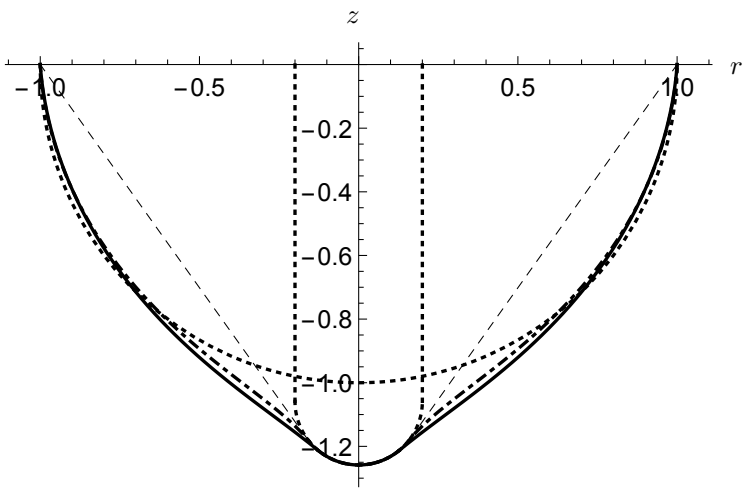} 
\caption{Comparison between the spherical indentation of a hemispherical surface modelled using the membrane model given in \citet{pearce2011} (dashed line), and both the approximate ($\xi=0$, dotdashed line) and full ($\xi=1$, solid line) shell models described above with $\beta=0.1$. In all cases $\rho = 0.2$ and the Mooney-Rivlin strain-energy function is used with $\alpha=0.1$. Dotted lines show the position of the indenter and the undeformed surface. 
}
\label{shellmembranecompare}
\end{figure}

For larger indentation depths it is possible to find solutions to the membrane equations, these have a cone-like shape as shown in Figure \ref{shellmembranecompare} where the straight lines reflect the average surface in the regions of compressive stress where the membrane is wrinkled. In comparison, the two shell theories show that the inclusion of the bending stiffness means that the shell requires energy to change the curvature from the reference state, ensuring that the deformation is localised rather than global.

\begin{figure}[!htb]
\psfrag{aa}{$r$}
\psfrag{bb}{$z$}
\includegraphics[width=0.5\textwidth,keepaspectratio]{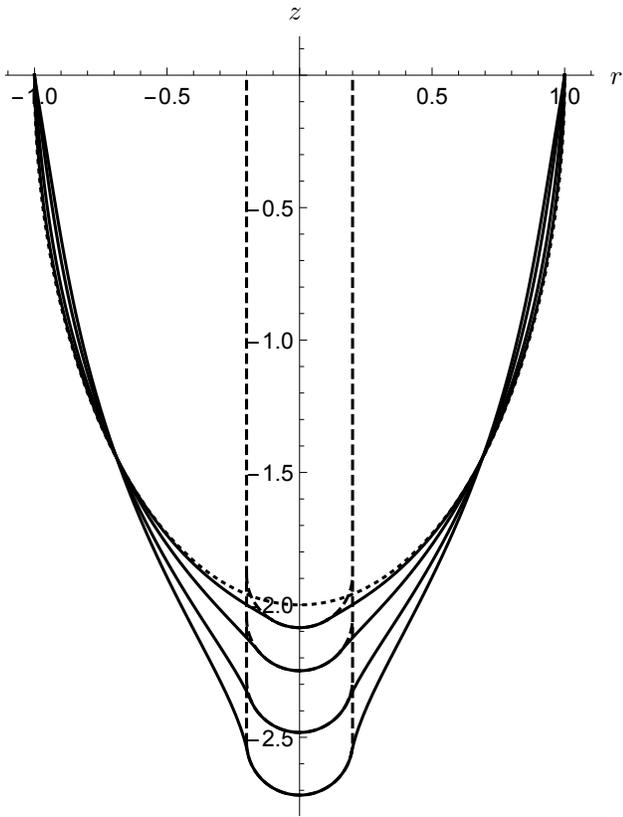} 
\caption{Profile of the deformed configuration for successive depths of indentation of a prolate spheroidal shell, $\gamma = 2$, by a spherical indenter with $\rho = 0.2$ and the Mooney-Rivlin model with $\alpha=0.1$. Dashed lines show the position of the indenter and the undeformed surface.}
\label{pro2RZindent}
\end{figure}

\begin{figure}[!htb]
\psfrag{aa}{$r$}
\psfrag{bb}{$z$}
\includegraphics[width=0.5\textwidth,keepaspectratio]{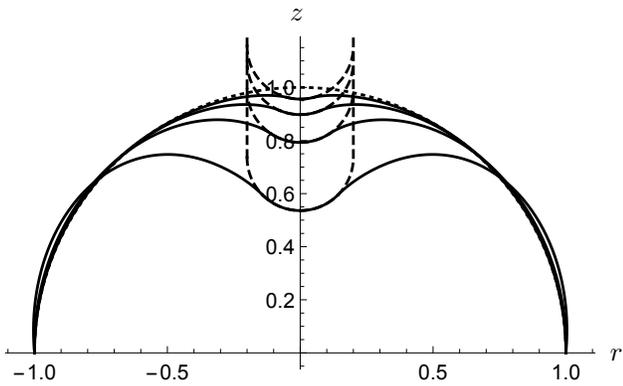} 
\caption{Profile of the deformed configuration for successive depths of indentation of a spherical shell from the concave side, $\gamma = -1$, by a spherical indenter with $\rho = 0.2$ and the Mooney-Rivlin model with $\alpha=0.1$. Dashed lines show the position of the indenter and the undeformed surface.}
\label{upsphRZindent}
\end{figure}

Two further examples are shown, a prolate spheroid indented from the concave side (Figure \ref{pro2RZindent}) and a sphere being indented from the convex side (Figure \ref{upsphRZindent}), illustrating how the shape of the deformed configuration changes with increasing depth of indentation. 


Experimentally, the readily measurable variables during an indentation test are the position and the total axial force exerted upon the indenter, in dimensional terms. We are therefore interested in non-dimensional force-displacement curves generated from solving the equations over a range of $\lambda_0$, which may then be used to estimate the elastic moduli of the sample by comparison with the experimental curves. 

In order to generate a force-displacement curve, we need to specify a number of parameters. These are the relative size and shape of the indenter ($\rho, A, B$), the shape of the undeformed shell ($R, Z$), the strain-energy function ($W$) and the bending-stretching ratio $\beta$. Here we will show the effect of varying these parameters, with the base-case parameters being a spherical shell and indenter with $\rho = 0.2, \beta = 0.01$, a Mooney-Rivlin strain-energy function with $\alpha=0.1$ and the full expansion $\xi=1$.

\begin{figure}[!htb]
\psfrag{aa}{$\delta$}
\psfrag{bb}{$F$}
\includegraphics[height=2.5in,keepaspectratio]{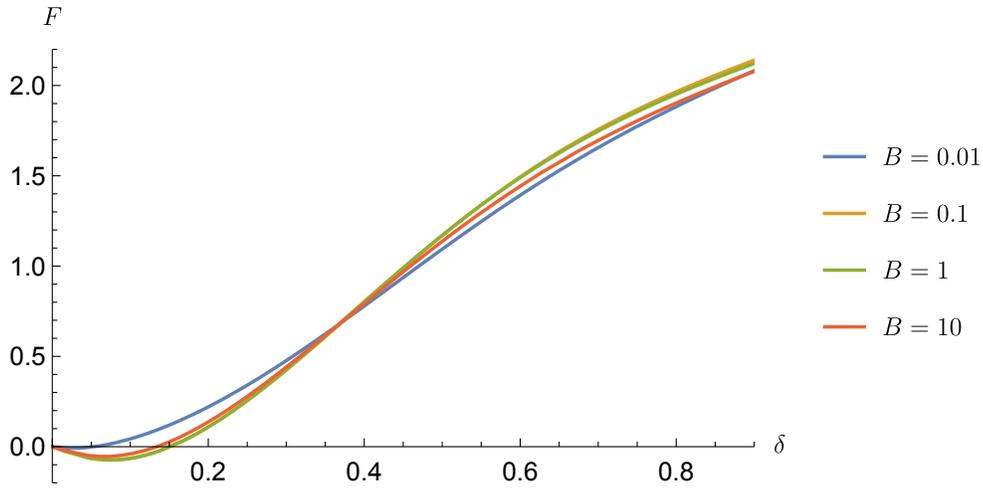} 
\caption{Changing the bending/stretching ratio $\beta$ with the approximate theory, $\xi=0$, has little impact on the force-displacement curve. All three curves give a negative force for small indentation distance.}
\label{Bvary}
\end{figure}

The approximate theory as defined $\xi=0$ in \rr{tensions}, works until the reference configuration of the shell becomes increasingly prolate spheroidal, at which point the neglected terms proportional to $M \kappa$ become significant. At this point, as $\delta\to0$ the two opposing terms in \rr{Flaw} both tend towards zero, but with the transverse shear term being slightly larger. This leads to a negative force $F$, even though the calculated shape looks appropriate, this can occur for spherical shells (depending on the other parameters) including for the parameters considered in Figure \ref{shellmembranecompare}. Figure \ref{Bvary} shows how these force-displacement curves are continuous, but with a small initial region where they go negative, where the curves go below the axis for $\delta<0.2$. Using smaller values of $B$ prevents this from happening for this set of parameters, but the issue reoccurs as $\gamma$ increases and the equations become increasingly numerically stiff as $B$ is reduced.

\begin{figure}[!htb]
\psfrag{aa}{$\delta$}
\psfrag{bb}{$F$}
\includegraphics[height=2.5in,keepaspectratio]{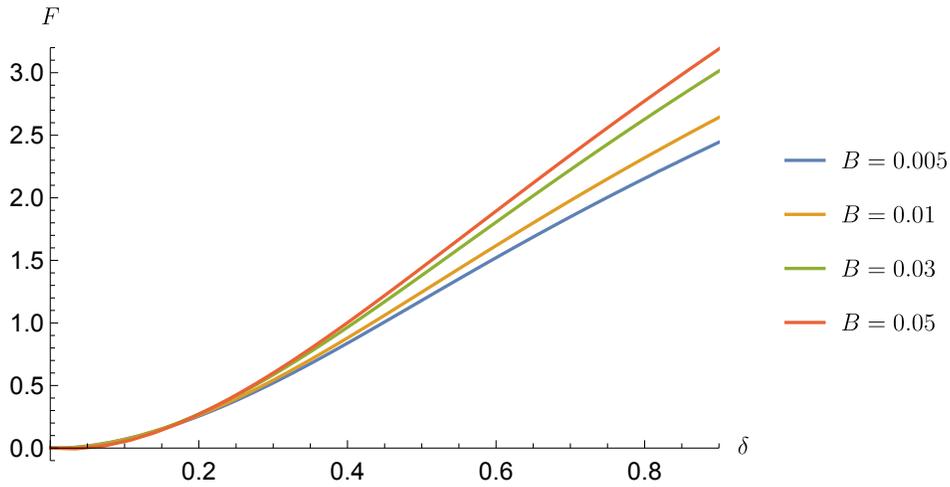} 
\caption{Changing the bending/stretching ratio $\beta$ with the full theory, $\xi=1$, increases the force required to indent the shell and removes the negative values of the force as seen for $\xi=0$.}
\label{Bvary2}
\end{figure}

Using the full theory, $\xi=1$, prevents this counter-intuitive behaviour (see Figure \ref{Bvary2}), and allows the calculation of a force-displacement curve which remains positive in the small indentation limit. The neglected $M \kappa$ terms become particularly relevant in this application because of the abrupt change in the curvature in the indenter region, where the shell is required to conform to the indenter. This generates large bending moments without a corresponding large stretch, meaning that the balance between the two terms in \rr{Flaw} is not masked by a large tensile stress resultant. We believe that this has not been noticed in the previous literature on the indentation of curved shells due to the fact that here we indent from the concave side rather than the convex side. We note that it is possible that the negative forces occur due to the surface not remaining in contact with the indenter throughout the contact region, but as it occurs in the limit of small $\delta$ (where $S_c$ also goes to zero) and from observing the resulting shapes this does not seem to be the case here.

As the aspect ratio of the spheroid becomes increasingly large (above $\gamma\approx2.5$ for $\rho =0.2$), the negative force reappears for very small indentation; it is not clear what other modifications are required to the theory to prevent this from happening here. Given that all shell theories involve decisions on which terms can be neglected, we suggest there there are some further neglected terms which are becoming relevant.

In addition to the negative forces as discussed earlier, when using the approximate theory, $\xi=0$, varying the bending to stretching ratio $\beta$ has only a marginal effect on the required force, even for extremely large values of $\beta$, as shown in Figure \ref{Bvary}. However, when we use the full theory, $\xi=1$, changing $\beta$ does have a significant effect on the force-indentation curve, as would be expected and is seen in Figure \ref{Bvary2}. 

\begin{figure}[!htb]
\psfrag{a}{$\delta$}
\psfrag{b}{$F$}
\includegraphics[height=2.5in,keepaspectratio]{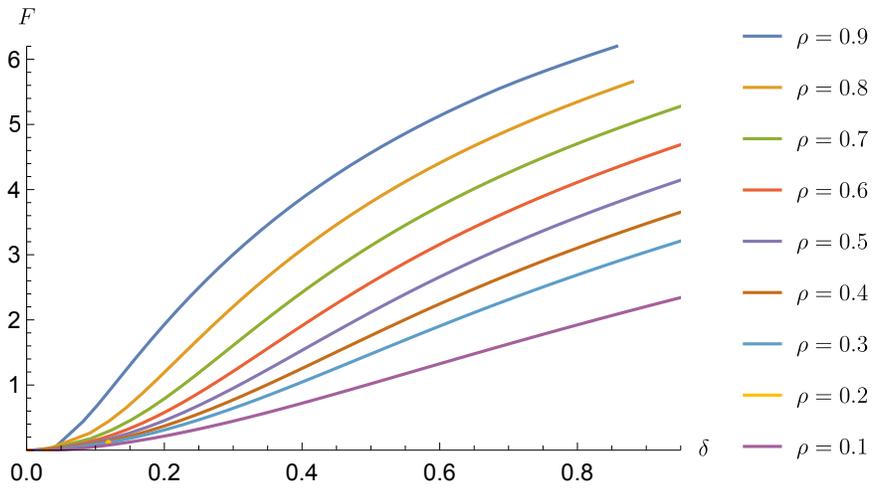} 
\caption{Effect of increasing the relative size of the indenter, $\rho$.}
\label{rhovary}
\end{figure}

As may be intuitively expected, increasing the relative size of the indenter has a strong effect on the total force required to achieve the same deformation, as may be seen in Figure \ref{rhovary}, but the shape of the curves are similar. Varying the size of the indenter allows for the collection of additional data for fitting purposes, which may be particularly useful in fitting the strain-energy function in the large-strain region.

\begin{figure}[!htb]
\includegraphics[height=2.5in,keepaspectratio]{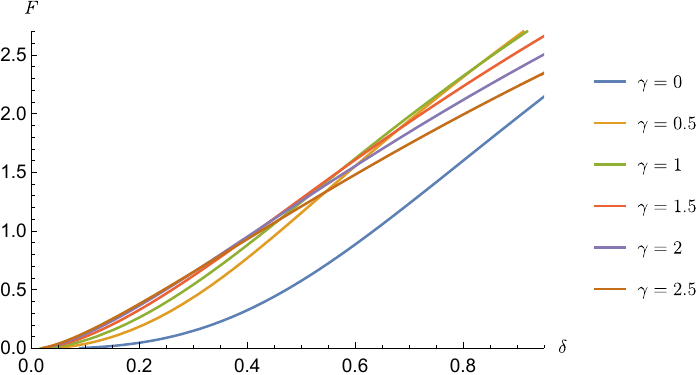} 
\caption{Effect of changing the undeformed shape of the shell, $R(S) = \sin S, Z(S) = - \gamma \cos S$.}
\label{RZvary}
\end{figure}

Varying the reference shape of the shell by making it more prolate  increases the initial steepness of the force-displacement curve, but this increase is not uniform across the depth of indentation for highly curved shells (Figure \ref{RZvary}). For these prolate spheroid cases with $\gamma>2$ here, further indentation does not require as much stretching as they are already highly curved, and therefore the force-displacement curve has a different shape.

\begin{figure}[!htb]
\psfrag{aa}{$\delta$}
\psfrag{bb}{$F$}
\includegraphics[height=2.5in,keepaspectratio]{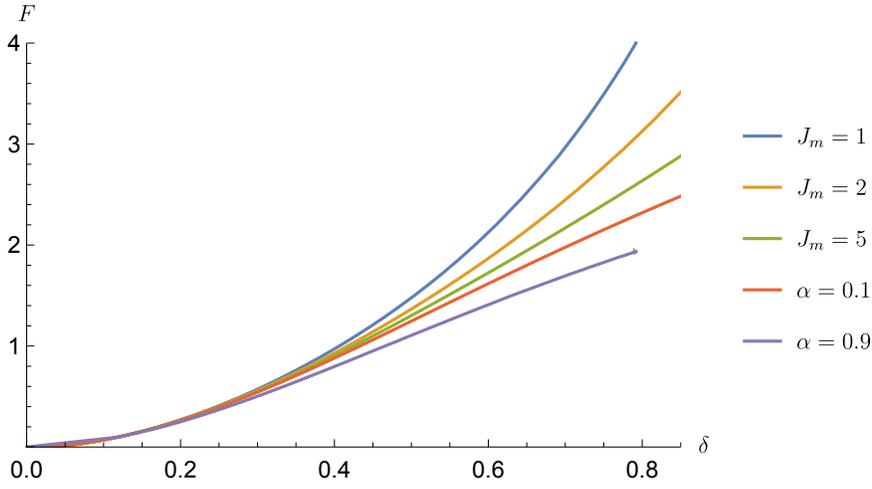} 
\caption{Effect of changing the strain-energy function, varying the parameters $\alpha$ and $J_m$ in the Mooney-Rivlin and Gent constitutive laws. At small-strains both strain-energy functions give the linear elasticity response but they diverge as the strains get larger.}
\label{Wvary}
\end{figure}

The choice of strain-energy function makes a significant difference only as the strains become moderate ($\delta>0.3$), but does has a noticeable effect there (Figure \ref{Wvary}). We note that the neo-Hookean strain-energy function ($\alpha=0$) has an unphysical limit at large strain, where increasing displacement does not require any increase in the applied force, as found in the membrane case \citep{pearce2011}. The Gent strain-energy function shows stiffening when the limiting parameter $J_m$ becomes small, as expected. We emphasize that the theory presented here can be used for any isotropic incompressible strain-energy function.

\begin{figure}[!htb]
\includegraphics[height=2.5in,keepaspectratio]{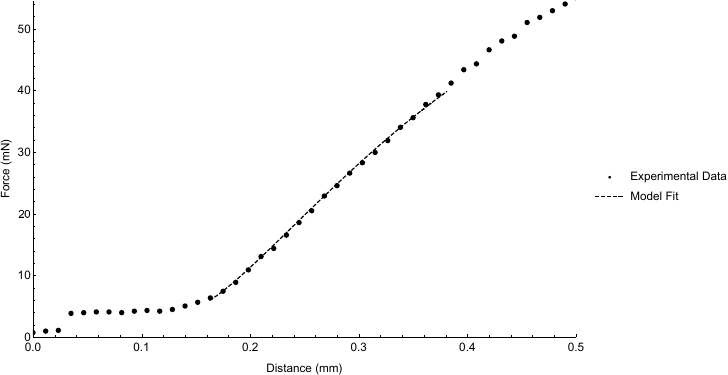} 
\caption{Example of fitting to experimental data, generated in a set-up similar to that in Figure \ref{indentimage} but using a smaller indenter. The dashed curve is from the model presented here, with a spherical indenter, a Mooney-Rivlin strain-energy function with $\alpha=0.1$, $\rho=1/3, \gamma=2.5, B=0.01$, and then fitted to the experimental curve with $\hat{\mu} \hat{H} = \SI{52}{\mega \pascal \micro \metre}$.}
\label{fitfigure}
\end{figure}

Finally, we show a preliminary fitting to a set of experimental data measured from a Lepidium endosperm, similar to the curve shown in Figure \ref{indentimage}. In the indentation shown in Figure \ref{indentimage}, the indenter is large enough to touch the side of the curved endosperm before it reaches the end, and the theory presented here needs modifying to accommodate that. We therefore fit to a curve from a smaller needle with a diameter of \SI{0.2}{\milli \meter}, and so we use $\rho=1/3, \gamma=2.5, B=0.01$ to generate the non-dimensional load-indentation curve . As shown in Figure \ref{fitfigure}, the experimental data is somewhat noisy at low force values and it is not entirely clear where the origin should be located, but a reasonable fit is achieved for $\hat{\mu} \hat{H} = \SI{52}{\micro \metre \mega \pascal}$. For an estimated endosperm thickness of \SI{50}{\micro \metre}, this gives a shear modulus of around $\SI{1}{\mega \pascal}$, a reasonable value for this kind of soft plant tissue.


\section{Conclusions}
We have developed a mathematical framework for the finite indentation of curved elastic shells, with the goal of characterising the shape-independent elastic properties from biological samples. The addition of the bending stiffness regularises the membrane equations, and allows curved surfaces to be considered. This may be used to compare between different shaped surfaces and is of particular use in a biological context, where the surface shape may vary between samples, or between regions of the same structure. In the specific case of the Lepidium endosperm, we will be able to extract the elastic properties of the different shaped sections covering the different parts of the seed, and see if the elastic modulus of the micropylar endosperm (the section that covers the root) becomes more pliant during germination. Growing tissues in particular are an application where being able to disentangle the shape from the elastic properties is important. Mathematical simulations of biological tissues often require knowledge of these elastic properties, so having an additional technique to measure them is useful in increasing the accuracy of such studies.

Within the theory as presented above, the only unknown parameters are the shear modulus, $\hat{\mu}$, the bending modulus, $\hat{B}$ (which may be connected by \rr{Bdef}), and the form of the strain-energy function, $W$, including any constants within. It is therefore appropriate first to extract the shear  modulus from the beginning of the curve, followed by the strain-energy function from large-strain data. 

One important finding is that when the shell becomes significantly curved, the approximate theory ($\xi=0$) which lacks the $M \kappa$ terms breaks down, predicting a negative force for small indentation, and the full theory ($\xi=1$) must be used. This effect is not dependent on the choice of constitutive law, and comes from the neglecting the contribution of the bending of the shell reference surface. We therefore recommend caution when using the approximate equations for curved shells, although they are commonly used. As the spheroid becomes increasingly prolate ($\gamma > 3$), these negative force terms reappear even when using the full theory $\xi=1$; it is not clear why this is occurring in this case.

The basis of this model could be extended in a number of directions. The indentation of shells from the convex side could be further considered, particularly in combination with an internal pressure $P$; this would change the pre-indentation shape of the shell (and hence change the boundary condition at $\lambda_\theta(L)$) but the governing equations are otherwise unchanged. Combined these two extensions allow for the modelling of the indentation of shells with an internal pressure, such as inflated balls, balloons or biological tissues. Other extensions to a non-uniform thickness, different initial configurations and boundary conditions (such as complete shells) or a bending energy that depends on further invariants of $\boldDelta$ are also possible.

The theory here restricts the shell to remain entirely in contact with the indenter at all times, this may not be the case if buckling occurs due to compressive stresses. The initial assumption of axisymmetry would not be valid in this case, and a post-buckling analysis would be required to look for other solutions.  This situation is most likely in the convex indentation, where higher compressive stresses appear \citep{vaziri2009}. Similarly, for indenters which are flatter than the shell they may not initially contact the shell at the tip; in this case two free boundaries will be present and the model will need to be adjusted to account for this.
 
There are also similarities between this indentation problem and the shaft-loaded blister test \citep{chopin2008, zhao2011,wan1999}, where the delamination of a flat elastic surface from a rigid substrate is measured by indentation, and hence the framework described here may also be useful for extending the blister-test analysis to non-flat surfaces. 


\section*{Data Accessibility}
This work does not include any data.

\section*{Competing Interests}
We have no competing interests.

\section*{Authors' contributions}
SPP conceived and implemented the mathematical model, interpreted the results and drafted the manuscript. JRK conceived the model, secured funding, interpreted the results and helped draft the manuscript. TS performed experiments and revised the manuscript. NME provided conceptual advice, discussed the results and revised the manuscript. GLM and MJH supervised the work, secured funding and provided the biological motivation. All authors gave final approval for the publication.

\section*{Acknowledgements}
SPP particularly thanks the many people with whom he has discussed the negative forces, and the anonymous reviewers for their helpful comments which have significantly improved the manuscript.

\section*{Funding Statement}
This work was initially funded by an ERANET-PG grant for the vSEED project (BBSRC grant BBG02488X1, DFG grant LE720/8-1). SPP thanks the Leverhulme Trust for the award of an Early Career Fellowship. JRK  acknowledges the support of the Royal Society and the Wolfson Foundation. GLM and TS acknowledge support from BBSRC grant BBM0005831.


\clearpage
\bibliographystyle{unsrtnat}
\bibliography{Indentref}

\end{document}